# Gas-grain model of carbon fractionation in dense molecular clouds

*Jean-Christophe Loison[1]\*, Valentine Wakelam[2], Pierre Gratier[2] and Kevin M. Hickson[1]*

\*Corresponding author: jean-christophe.loison@u-bordeaux.fr

[1] Institut des Sciences Moléculaires (ISM), CNRS, Univ. Bordeaux, 351 cours de la Libération, 33400, Talence, France
[2] Laboratoire d'astrophysique de Bordeaux, Univ. Bordeaux, CNRS, B18N, allée Geoffroy Saint-Hilaire, 33615 Pessac, France.

Carbon containing molecules in cold molecular clouds show various levels of isotopic fractionation through multiple observations. To understand such effects, we have developed a new gas-grain chemical model with updated $^{13}$C fractionation reactions (also including the corresponding reactions for $^{15}$N, $^{18}$O and $^{34}$S). For chemical ages typical of dense clouds, our nominal model leads to two $^{13}$C reservoirs: CO and the species that derive from CO, mainly s-CO and s-CH$_3$OH, as well as C$_3$ in the gas phase. The nominal model leads to strong enrichment in C$_3$, c-C$_3$H$_2$ and C$_2$H in contradiction with observations. When C$_3$ reacts with oxygen atoms the global agreement between the various observations and the simulations is rather good showing variable $^{13}$C fractionation levels which are specific to each species. Alternatively, hydrogen atom reactions lead to notable relative $^{13}$C fractionation effects for the two non-equivalent isotopologues of C$_2$H, c-C$_3$H$_2$ and C$_2$S. As there are several important fractionation reactions, some carbon bearing species are enriched in $^{13}$C, particularly CO, depleting atomic $^{13}$C in the gas phase. This induces a $^{13}$C depletion in CH$_4$ formed on grain surfaces, an effect that is not observed in the CH$_4$ in the solar system, in particular on Titan. This seems to indicate a transformation of matter between the collapse of the molecular clouds, leading to the formation of the protostellar disc, and the formation of the planets. Or it means that the atomic carbon sticking to the grains reacts with the species already on the grains giving very little CH$_4$.

# 1. Introduction

The study of molecular isotopologues provides an important opportunity to understand the transformation of interstellar matter from gas and dust to the eventual formation of planetary systems. In order to gain a better insight into the various processes leading to isotope fractionation in N-bearing, O-bearing and S-bearing molecules, we have performed a series of investigations related to the chemistry occurring in dense molecular clouds with a particular emphasis on $^{15}$N (Loison *et al.* 2018), $^{18}$O (Loison *et al.* 2019) and $^{34}$S (Loison et al. 2019) reactions. In these regions, low temperatures are thought to play an important role in the enrichment process, as zero point energy (ZPE) differences between the reagents and products of a fractionation reaction favor the formation of molecules containing heavy isotopes. During these earlier studies, we demonstrated that the observed $^{14}$N/$^{15}$N ratios in nitrogen containing molecules could not be explained by the differing reactivity of these species. In contrast, the chemistry was shown to induce a significant enrichment in $S^{18}O$, $SO^{18}O$, $N^{18}O$ and $O^{18}O$ molecules (with a low corresponding enrichment in $^{34}$S), allowing the chemical age of dense interstellar clouds to be better constrained. Here, we describe a study of $^{13}$C fractionation in interstellar molecules that have the characteristic of involving a large number of observed species which also present diverse fractionation levels. While carbon-bearing molecules display variable, and sometimes large, isotopic fractionation levels in diffuse (Liszt 2007, Ritchey *et al.* 2011) and dense molecular clouds (see Table 1), the derived $^{12}$C/$^{13}$C ratios in the Solar System are remarkably constant, with values close 90 in solar CO (Ayres *et al.* 2013), in Titan's atmosphere (Courtin *et al.* 2011, Gurwell 2004, Jennings *et al.* 2008, Jennings *et al.* 2009, Mandt *et al.* 2012) and in comets (Bockelée-Morvan *et al.* 2015), all three being close to the telluric value. It should be noted that the solar system $^{12}$C/$^{13}$C ratio close to 90 is different from the local ISM value of $^{12}$C/$^{13}$C = 68 ± 15 suggested by (Milam *et al.* 2005) as the sun may have migrated from its original birthplace closer to the Galactic Centre (Romano *et al.* 2017). The first study on $^{13}$C fractionation was carried out by Langer *et al.* (1984) showing the importance of the $^{13}C^+ + CO \rightarrow C^+ + {}^{13}CO$ reaction. Roueff *et al.* (2015) expanded on this earlier work by showing that it was necessary to include other processes such as the $^{13}C + CN \rightarrow C + {}^{13}CN$ reaction. Very recently, Colzi *et al.* (2020) have performed the first gas-grain model for $^{13}$C fractionation adding few $^{13}$C fractionation reactions from Roueff et al. (2015). In their study, Colzi et al. (2020) have considered all carbons of a given species as equivalents, e.g. they do not differentiate $^{13}$CCH and C$^{13}$CH, which induces several biases in the results and does not allow a realistic comparison with most of the observations. We have extended these $^{13}$C fractionation studies considering the various positions for $^{13}$C as specific species (e.g. $^{13}$CCH and $C^{13}CH$ are different species) and including various new chemical reactions such as those involving secondary carbon reservoirs such as $C_3$ and HCN. We limited our study to dense clouds where photodissociation induced fractionation is minor. In diffuse molecular clouds and circumstellar disks, $^{12}CO$ self-shielding provides a mechanism for selective photodissociation of $^{13}CO$, thereby increasing the $^{12}CO/^{13}CO$ ratio (Visser *et al.* 2009).

Table 1: Observations for $^{13}C$ in dense molecular clouds.

| species | ratio | Cloud | references |
|---|---|---|---|
| $C^{18}O/^{13}C^{18}O$ | 70(20) | L1527 | (Yoshida *et al.* 2019) |
| $C^{17}O/^{13}C^{17}O$ | 42(13) | L483 | (Agúndez *et al.* 2019) |
| $HCO^+/H^{13}CO^+$ | 49(14) | TMC1 | (Turner 2001) |
| $CCH/^{13}CCH$ | > 250 | TMC1 | (Sakai *et al.* 2010) |
| | > 250 | TMC1 | (Liszt & Ziurys 2012) |
| | 210(60) | L1527 | (Yoshida et al. 2019) |
| | >162 | L483 | (Agúndez et al. 2019) |
| $CCH/C^{13}CH$ | > 170 | TMC1 | (Sakai et al. 2010) |
| | > 170 | TMC1 | (Liszt & Ziurys 2012) |
| | 140(40) | L1527 | (Yoshida et al. 2019) |
| | >70 | L483 | (Agúndez et al. 2019) |
| $c-C_3H_2/c-CC^{13}CH_2$ | 61(11) | L1527 | (Yoshida *et al.* 2015) |
| | 41(8) | L1527 | (Yoshida et al. 2015, Yoshida et al. 2019) |
| | 53(16) | L483 | (Agúndez et al. 2019) |
| $c-C_3H_2/c-^{13}CCCH_2$ | 310(80) | L1527 | (Yoshida et al. 2015) |
| | 200(30) | L1527 | (Yoshida et al. 2019) |
| | 458(138) | L483 | (Agúndez et al. 2019) |
| $CH_3CCH/^{13}CH_3CCH$ | 60(18) | L483 | (Agúndez et al. 2019) |
| $CH_3CCH/CH_3^{13}CCH$ | 53(16) | L483 | (Agúndez et al. 2019) |
| $CH_3CCH/CH_3C^{13}CH$ | 58(17) | L483 | (Agúndez et al. 2019) |
| $CN/^{13}CN$ | 50(20)* | B1 | (Daniel *et al.* 2013) |
| | 44(8) | OMC-2 | (Kahane *et al.* 2018) |
| | 61(17) | L1527 | (Yoshida et al. 2019) |
| $HCN/H^{13}CN$ | 30(7)** | B1 | (Daniel et al. 2013) |
| | 45(3)* | L1498 | (Magalhães *et al.* 2018) |
| $HNC/HN^{13}C$ | 43-72** | TMC1 | (Liszt & Ziurys 2012) |
| | 20(5)** | B1 | (Daniel et al. 2013) |
| $HC_3N/H^{13}CCCN$ | 79(11) | TMC1 | (Takano *et al.* 1998, Liszt & Ziurys 2012) |
| | 132(10) | TMC1 | (Gratier *et al.* 2016) |
| | 57(7) | OMC-2 | (Kahane et al. 2018) |
| | 86(16) | L1527 | (Araki *et al.* 2016) |
| | 85(22) | L1527 | (Yoshida et al. 2019) |
| | 91(27) | L483 | (Agúndez et al. 2019) |
| $HC_3N/HC^{13}CCN$ | 75(10) | TMC1 | (Takano et al. 1998, Liszt & Ziurys 2012) |
| | 129(10) | TMC1 | (Gratier et al. 2016) |
| | 59(11) | OMC-2 | (Kahane et al. 2018) |
| | 85(17) | L1527 | (Araki et al. 2016) |
| | 51(7) | L1527 | (Yoshida et al. 2019) |
| | 93(28) | L483 | (Agúndez et al. 2019) |
| $HC_3N/HCC^{13}CN$ | 55(7) | TMC1 | (Takano et al. 1998, Liszt & Ziurys 2012) |
| | 79(8) | TMC1 | (Gratier et al. 2016) |
| | 46(6) | OMC-2 | (Kahane et al. 2018) |
| | 64(2) | L1527 | (Araki et al. 2016) |
| | 49(15) | L1527 | (Yoshida et al. 2019) |
| | 79(24) | L483 | (Agúndez et al. 2019) |
| $HC_7N/HC_6^{13}CN$ | 85(35) | TMC1 | (Burkhardt *et al.* 2018) |
| $HC_7N/HC_3^{13}CC_3N$ | 110(16) | TMC1 | (Cordiner *et al.* 2017) |
| | 90(33) | TMC1 | (Burkhardt et al. 2018) |
| $HC_7N/HC_4^{13}CC_2N$ | 96(11) | TMC1 | (Cordiner et al. 2017) |
| | 67(28) | TMC1 | (Burkhardt et al. 2018) |
| $CS/^{13}CS$ | 68(5) | TMC1 | (Liszt & Ziurys 2012) using $CS/C^{34}S = 22.7$ |
| | 58(18) | L483 | (Agúndez et al. 2019) using $CS/C^{34}S = 22.5$ |
| $H_2CS/H_2^{13}CS$ | 79(26) | TMC1 | (Liszt & Ziurys 2012) using $H_2CS/H_2C^{34}S = 22.7$ |
| | 113(34) | L483 | (Agúndez et al. 2019) |
| $CCS/^{13}CCS$ | 230(130) | TMC1 | (Sakai *et al.* 2007) |
| $CCS/C^{13}CS$ | 54(2) | TMC1 | (Sakai et al. 2007) |
| | 28(8) | L483 | (Agúndez et al. 2019) |

\*: main isotopologue shows opacity
\*\*: main isotopologue shows very strong opacity

In Section 2, we present the chemical model, describing the important isotopic exchange reactions and the main changes from Roueff et al. (2015). The results of the nominal model, the various tests used to examine the effects of certain reactions and the comparison with dense molecular cloud observations and the comparison with the results from Colzi et al. (2020) are shown in Section 3. Our conclusions are presented in Section 4.

## 2 The chemical model
### 2.1 Model description

Our model is the same as in our recent studies of $^{15}N$, $^{18}O$ and $^{34}S$ fractionation (Loison et al. 2018, Loison et al. 2019) based on the chemical model Nautilus in its 3-phase version from Ruaud *et al.* (2016) using kida.uva.2014[1] (Wakelam *et al.* 2015) as the basis for the reaction network. The new network is limited to carbon skeletons up to $C_3H_x$ (x = 0-2) and $C_3H_x^+$ (x = 0-3), to limit the number of reactions when considering all the $^{15}N$ and $^{13}C$ species. Species with a carbon $^{13}C$ and having several non-equivalent carbons ($C_2H$, $C_2S$, $C_3$, c-$C_3H$, l-$C_3H$, ...) are treated as different species. We did not consider multiple fractionations (with two or more $^{13}C$ or with one $^{13}C$ and one $^{15}N$, $^{18}O$, $^{34}S$) as they involve very minor species. We did not include either $CH_3CCH$ or $C_3H_6$ because these species cannot be efficiently produced by current gas-grain models and are not strongly linked to species present in our reduced network. The low production of $CH_3CCH$ or $C_3H_6$ by current gas-grain models is due to the fact that the ionic pathway via the $C_3H_3^+ + H_2$ reaction is supposed to be inefficient according to Lin *et al.* (2013a), and the desorption mechanisms are not efficient enough either despite the relatively large amount of $CH_3CCH$ and $C_3H_6$ formed on the surface of grains (formed by successive hydrogenation reactions of s-$C_3$) considering the usual desorption processes. The current network includes 4096 gas-phase reactions and 5586 grain reactions. We have checked than the new network reproduces the abundances of the complete network for the main species studied here. We have also introduced the $^{15}$N, $^{13}$C, $^{18}$O and $^{34}$S exchange reactions using an updated version of the network presented in Roueff et al. (2015) and Colzi et al. (2020) including various new exchange reactions, the nitrogen network being presented in our recent study (Loison et al. 2018) and the oxygen and sulfur exchange reactions in (Loison et al. 2019). The surface network is similar to the one developed by Ruaud *et al.* (2015) with additional updates from Wakelam *et al.* (2017).

---
[1] http://kida.obs.u-bordeaux1.fr/models

The chemical composition of the gas-phase and the grain surfaces is computed as a function of time. The gas and dust temperatures are equal to 10 K, the total $H_2$ density is equal to $2\times10^4$ cm$^{-3}$ (various runs have been performed with a total H density between $1\times10^4$ cm$^{-3}$ and $2\times10^5$ cm$^{-3}$, although the density does not appear to be a critical factor for fractionation). The cosmic-ray ionization rate is equal to $1.3\times10^{-17}$ s$^{-1}$ and the total visual extinction is set equal to 10. All elements are assumed to be initially in atomic form (elements with an ionization potential below the maximum energy of ambient UV photons (13.6 eV, the ionization energy of H atoms) are initially in a singly ionized state, i.e., C, S and Fe), except for hydrogen, which is entirely molecular. The initial abundances are similar to those of Table 1 of Hincelin *et al.* (2011), the C/O elemental ratio being equal to 0.7 in this study. The grains are considered to be spherical with a 0.1μm radius, a 3 g.cm$^{-3}$ density and about $10^6$ surface sites, all chemically active. The dust to gas mass ratio is set to 0.01.

## 2.2 Update of the chemistry ($^{13}$C exchange reactions)

In interstellar media, fractionation reactions are driven by zero point energy (ZPE) differences that favor one direction (reverse or forward) for barrierless exchange reactions at low temperature. Following our previous study of $^{18}$O fractionation (Loison et al. 2019) based on Henchman and Paulson (1989) we consider for the various fractionation reactions (forward ($k_f$) corresponds to the exothermic pathway and reverse ($k_r$) the opposite pathway, see Table 2 for the units):

$k_f$ (T) = $\alpha \times (T/300)^\beta \times f(B,M)/(f(B,M) + \exp(\Delta E/kT))$,

$k_r$ (T) = $\alpha \times (T/300)^\beta \times \exp(\Delta E/kT)/(f(B,M) + \exp(\Delta E/kT))$

with $\alpha$ and $\beta$ given by capture theory, $f(B,M)$ is related to the symmetry of the system (Terzieva & Herbst 2000) and $\Delta E$ = exothermicity of the reactions (see Table 2) and T is the temperature.

We have extensively updated the exchange reactions from Roueff et al. (2015) and Colzi et al. (2020), introducing 10 new fractionation reactions. We highlight the role of the barrierless $^{13}C + C_3$ (number 8 of Table 2) reaction (Wakelam *et al.* 2009), as Colzi et al. (2020), but also of the $C + {}^{13}CCC$ reaction (number 9 of Table 2), which play a crucial role if $C_3$ is present at high abundance levels (specifically in the case where $C_3$ is unreactive with atomic oxygen). The $^{13}C$ exchange reactions are listed in Table 2 (and are cited in the text by their number in the same table). It should be noted that there are considerably more exchange reactions for $^{13}C$ than for $^{14}N$, $^{18}O$ or $^{34}S$ isotopes.

Exchange reactions are efficient only in the gas-phase because the addition channel is always favored in ice surface reactions (for example s-$^{13}C$ + s-$C_2$ → s-$^{13}CCC$ only, s- meaning on grain, with no isotopic exchange). However, diffusion and tunneling are mass dependent and are then not strictly equivalent for the various isotopologues. These differences are included in our model but have only a very small effect on $^{13}C$ fractionation.

For reactions involving $C_3$, there are two possibilities to incorporate a $^{13}C$ atom in the terminal position. Then the elemental $C_3/{}^{13}CCC$ ratio is equal to 34 instead of 68.

For electronic Dissociative Recombination (DR) branching ratios, considering the large exothermicity of such processes, we always consider scrambling of the $^{13}C$ distribution among the carbon skeleton. For example, the $^{13}CCH^+ + e^-$ and $C^{13}CH^+ + e^-$ DR reactions lead to equal amounts of CH and $^{13}CH$.

For bimolecular reactions we used existing theoretical calculations to deduce whether or not there was mixing of the carbon structure according to the possibilities of isomerization of the transient species. It should be noted that this possibility of scrambling does not affect the overall fractionation of the species but simply the relative fractionation of non-equivalent carbon atoms for a given species.

**Table 2**. Review of isotopic exchange reactions.
$k_f = \alpha \times (T/300)^\beta \times f(B,M)/(f(B,M) + \exp(\Delta E/kT))$ (left to right),
$k_r = \alpha \times (T/300)^\beta \times \exp(\Delta E/kT)/(f(B,M) + \exp(\Delta E/kT))$ (right to left),
$\alpha$ in $cm^3.molecule^{-1}.s^{-1}$, $\beta$ without unit.

| | Reaction | | α | β | ΔE(K) | f(B,M) | reference |
|---|---|---|---|---|---|---|---|
| 1. | $^{13}C^+ + C_2$ | → $C^+ + {}^{13}CC$ | 1.86e-9 | 0 | -26.4 | 2 | This work. Low uncertainty for this reaction. See also Colzi et al. (2020). |
| 2. | $^{13}C^+ + C_3$ | → $C^+ + {}^{13}CCC$ | 1.80e-9 | 0 | -28.0 | 2 | This work. Low uncertainty for this reaction. See also Colzi et al. (2020). |
| 3. | $C^+ + {}^{13}CCC$ | → $C^+ + C^{13}CC$ | 1.80e-9 | 0 | -15.0 | 1 | This work. Low uncertainty for this reaction. |
| 4. | $^{13}C^+ + CN$ | → $C^+ + {}^{13}CN$ | 3.82e-9 | -0.4 | -31.1 | 1 | (Roueff et al. 2015). Low uncertainty for this reaction. |
| 5. | $^{13}C^+ + CO$ | → $C^+ + {}^{13}CO$ | 4.6e-10 | -0.26 | -34.8 | 1 | (Watson *et al.* 1976, Liszt & Ziurys 2012). Low uncertainty for this reaction. |

| # | Reaction | | Rate | α | ΔE | | Notes |
|---|---|---|---|---|---|---|---|
| 6. | $^{13}C^+ + CS$ | $\to C^+ + ^{13}CS$ $\to S^+ + ^{13}CC$ | 2.0e-9 "0" | -0.4 | -26.3 +49 | 1 | This work, half of the capture rate constant by comparison with $C^+ + CO$. Low uncertainty for this reaction. Colzi et al. (2020). |
| 7. | $^{13}C + C_2$ | $\to C + ^{13}CC$ | 3.0e-10 | 0 | -26.4 | 2 | (Roueff et al. 2015). Low uncertainty for this reaction. |
| 8. | $^{13}C + C_3$ | $\to C + ^{13}CCC$ | 3.0e-10 | 0 | -28.0 | 2 | This work based on (Wakelam et al. 2009). Low uncertainty for this reaction. Colzi et al. (2020). |
| 9. | $C + ^{13}CCC$ | $\to C + C^{13}CC$ | 3.0e-10 | 0 | -15.0 | 1 | This work based on (Wakelam et al. 2009). Low uncertainty for this reaction. |
| 10. | $^{13}C + CN$ | $\to C + ^{13}CN$ | 3.0e-10 | 0 | -31.1 | 1 | (Roueff et al. 2015). Low uncertainty for this reaction. |
| 11. | $^{13}C + HCN$ | $\to C + H^{13}CN$ | 2.0e-10 | 0 | -48.0 | 1 | This work and (Loison & Hickson 2015). Relatively low uncertainty for this reaction (see text). |
| 12. | $^{13}C + HNC$ | $\to C + HN^{13}C$ $\to ^{13}C + HCN$ $\to C + H^{13}CN$ | 3.0e-11 6.0e-11 low | 0 | -33 | 1 | This work and (Loison & Hickson 2015). Relatively large uncertainty on branching ratio for this reaction (see text). |
| 13. | $^{13}C + HCNH^+$ | $\to C + H^{13}CNH^+$ | 1.0e-9 | 0 | -50.0 | 1 | This work. Intermediate uncertainty for this reaction (see text). |
| 14. | $^{13}C + HC_3N$ | $\to C + H^{13}CCCN$ $\to C + HC^{13}CCN$ $\to C + HCC^{13}CN$ $\to H + ^{13}CCCCN$ $\to H + C^{13}CCCN$ $\to H + CC^{13}CCN$ $\to H + CCC^{13}CN$ $\to HCN + ^{13}CCC$ $\to HCN + C^{13}CC$ $\to H^{13}CN + C_3$ | 2.0e-11 2.0e-11 2.0e-11 0 0 0 0 2.0e-11 1.0e-11 1.0e-11 | 0 0 0 | -48.3 -57.4 -60.0 | | The $C + HC_3N$ have been studied theoretically by Li et al. (2006). Isotopic exchanges involve multiple intermediate and are in competition with back dissociation, H + $C_4N$ production ($C_4N$ not considered in the current isotopic network) and $HCN + C_3$ through spin-orbit coupling. There are large uncertainties for this reaction (see text) |
| 15. | $^{13}C + CS$ | $\to C + ^{13}CS$ | 2.0e-10 | 0 | -26.3 | 1 | This work (see text). Low uncertainty for this reaction. |
| 16. | $H + ^{13}CCH$ | $\to H + C^{13}CH$ | 2.0e-10 | 0 | -8.1 | 1 | This work based on Harding et al. (2005), ΔE in agreement with Furuya et al. (2011). Low uncertainty for this reaction. |
| 17. | $H + c-^{13}C<CHCH$ | $\to H + c-C<C^{13}HCH$ | 2.0e-10 | 0 | -26.0 | 1 | This work (see text). Large uncertainties for this reaction which may show a barrier in the entrance valley. The c-$^{13}C<CHCH$ notation means that carbon $^{13}C$ is the one without a hydrogen atom which are described in the Appendix. |
| 18. | $H + ^{13}CCS$ | $\to H + C^{13}CS$ | 4.0e-11 | 0 | -18.0 | 1 | This work. Furuya et al. (2011) gives ΔE = -17.4 K. Low uncertainty for this reaction. |
| 19. | $HCNH^+ + H^{13}CN$ | $\to H^{13}CNH^+ + HCN$ | 2.0e-9 | -0.5 | -2.9 | 1 | This work based on Cotton et al. (2013). Low uncertainty for this reaction. |
| 20. | $HCNH^+ + HN^{13}C$ | $\to H^{13}CNH^+ + HCN$ $\to HCNH^+ + H^{13}CN$ | 1.0e-9 1.0e-9 | -0.5 -0.5 | 0 0 | | This work based on Cotton et al. (2013). Relatively large uncertainty for the $HCN/H^{13}CN$ branching ratio for this reaction. |
| 21. | $^{13}CO + HCO^+$ | $\to H^{13}CO^+ + CO$ | 2.6e-10 | -0.4 | -17.4 | 1 | (Smith & Adams 1980). Low uncertainty for this reaction. |

Among the various $^{13}C$ exchange reactions, the reaction 5 of Table 2 have been studied previously in detail (Watson et al. 1976, Liszt & Ziurys 2012). Some others have been estimated in Roueff et al. (2015) and in Colzi et al. (2020) based on the barrierless nature of radical-radical reactions. We extend this work to various new fractionation reactions, namely number 9, 11, 12, 13, 14, 15, 16, 17, 18, 19, and 20 of Table 2 (it should be noted that we use our own calculations for the reactions published in (Colzi et al. 2020), which leads to very small differences in the rate constant). All the reactions presented in Table 2 are likely to occur in the absence of a barrier based on their radical character, while certain processes have been studied theoretically in detail as noted in the table. Although the absence of an activation barrier is a requirement for effective isotope enrichment at interstellar temperatures, there must also be a reaction pathway that allows exchange. The possibility of exchange is obvious for some reactions such as 1, 7 or 16 it is much less obvious for some others such as 11, 12, 13, 14, 15, 17, 18 and requires theoretical calculations carried out in this work and presented in Appendix or from previous work already published and referenced in the Table 2.

As noted in the introduction, photodissociation of CO can play an important role in the diffuse phase prior to the formation of dense molecular clouds as $^{12}$CO self-shielding leads to selective photodissociation of $^{13}$CO and an increase in the $^{12}$CO/$^{13}$CO ratio (Visser et al. 2009). This effect also increases the abundance of $^{13}$C$^+$. Even if it is not the subject of this study we looked at the influence of the initial conditions starting either from C$^+$ and $^{13}$C$^+$ only or with carbon initially present as CO with a variable $^{12}$CO/$^{13}$CO ratio to simulate the fact that $^{13}$CO is more easily photo-dissociated than $^{12}$CO in diffuse clouds due to the self-shielding effect of $^{12}$CO. In the latter case, the excess of $^{13}$C$^+$ is rapidly consumed by reaction 5 and transformed into $^{13}$CO. Then the abundances of the different species, including $^{13}$C species, are similar at the time when C$^+$ is transformed into neutral C (around a few 10$^3$ years) as long as the initial carbon is not entirely in the form of CO. So, unlike nitrogen chemistry where the preferential photodissociation of $^{15}$NN in the diffuse phase prior to the formation of dense molecular clouds may lead to an overabundance of $^{15}$N in the gas phase of dense molecular clouds (Furuya & Aikawa 2018), the photodissociation of CO in the diffuse phase prior to the formation of dense molecular clouds does not seem to have a large influence on carbon fractionation in dense molecular clouds even if a dedicated study should be carried out in the future.

## 3. Results
### 3.1 General results on the $^{12}$C/$^{13}$C ratios
The time dependence of the $^{12}$C/$^{13}$C ratios given by our nominal model for various species are shown in Figure 1.

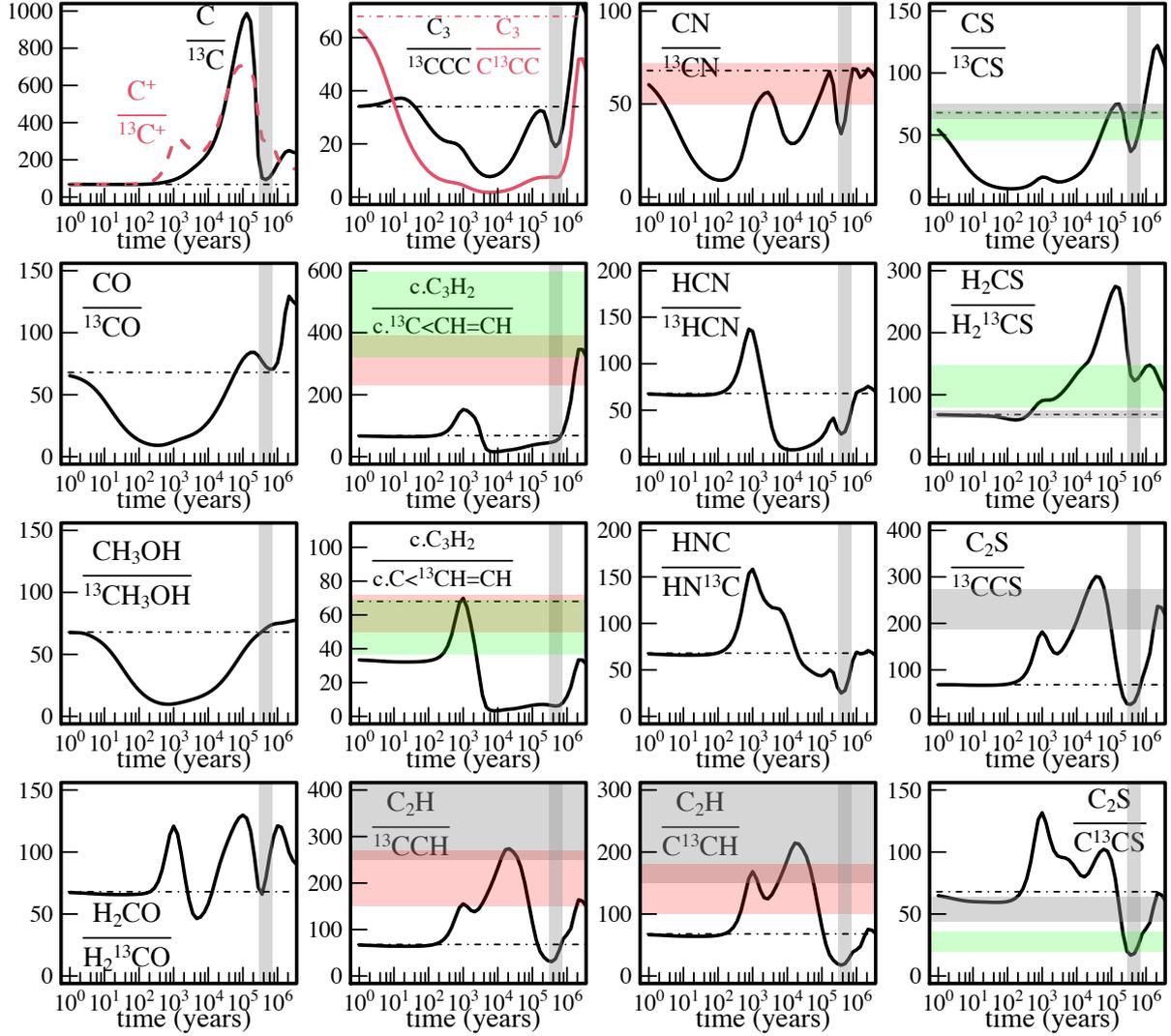

**Figure 1:** Gas phase species of $^{12}C/^{13}C$ ratio abundances of various species as a function of time predicted by our model (n(H$_2$) = 2×10$^4$ cm$^{-3}$, T = 10K) with a rate constant for the O + C$_3$ reaction equal to 1×10$^{-14}$ cm$^3$.molecule$^{-1}$.s$^{-1}$. The dashed horizontal lines represent the "cosmological" $^{12}C/^{13}C$ ratio equal to 68 (34 for C$_3$/$^{13}$CCC and c-C$_3$H$_2$/c-C$^{13}$CHCH taking into account the statistical factor of 2 for the two equivalent carbon atoms). The observations in dense molecular clouds are reported in horizontal rectangles (including the uncertainties), in light grey for TMC-1(CP), in light red for L1527, in light blue for L1498 and in light green for L483 (see Table 1 for the references). The vertical grey rectangles represent the values given by the most probable chemical age for TMC-1(CP) and L483 given by the better agreement between calculations and observations for key species.

Our model leads to high $^{13}C$ fractionation after few 10$^4$ years for most of the species because a large part of the $^{13}C$ is locked into $^{13}CO$ (and species derived from CO i.e. s-CH$_3$OH and s-CO$_2$ on grains) but also C$^{13}$CC (C$_3$ reaches a large abundance, up to 10% of total elemental carbon) with a minor contribution from H$^{13}$CN. It should be noted that fractionation induced by the slight difference in sticking rate between $^{12}C$-atoms and $^{13}C$-atoms is negligible in the case of carbon as chemical fractionation involves much larger fluxes than the difference induced by depletion. The mechanisms at work in $^{13}C$ fractionation arise from the efficient enrichment reactions with the main carbon reservoir (CO) and the secondary reservoirs C$_3$ and HCN. The reactions 5, 8 and 9 are efficient at early times when the $C^+$ abundance is high, and the reactions 8, 9 and 11 are important at later times. It should be noted that to be efficient, the exchange reactions should be barrierless with an exothermicity significantly above the dense molecular cloud temperature ( > 10 K). Additionally, as these processes need to involve large fluxes to induce important fractionation effects, they should involve species with large gas phase abundances or fast reaction rates. At early times, around $10^3$ years, the reaction 5 leads to notable enrichment in CO and, as CO is already quite abundant, this leads to $^{13}C$ depletion in the majority of other carbon containing species. For intermediate time, between $10^4$ years and $10^5$ years, the fractionation reactions lead to enrichment in $C^{13}CC$ and $H^{13}CN$. For the longer times characteristic of dense cloud ages (between $3\times10^5$ and $7\times10^5$ years) the fractionation reactions lead to enrichment mainly in $C^{13}CC$. The estimation of the dense cloud ages is given by the best agreement between calculations and observations for key species given by the so-called distance of disagreement (Wakelam *et al.* 2006), the chemical age being very likely not the same for the different molecular clouds observed, being function of the density and structure of the observed clouds, and requires a large number of observed species to be determined such as for TMC-1(CP) or L483), The amount of $^{13}C$ in $C_3$ is so large that carbon atom in the gas phase becomes highly depleted and species not linked to $C_3$, such as $H_2CS$, are depleted in $^{13}C$. Even CO is slightly depleted in $^{13}C$ due to the very large enrichment of $C^{13}CC$, but also due to the fact that at late time the $C^+$ abundance is low and then the $^{13}C^+ + CO$ reaction is inefficient preventing the CO produced at late time to be enriched in $^{13}C$. Species whose observed $^{12}C/^{13}C$ ratio is well reproduced by the simulations are those for which fractionation is low (CO, $HCO^+$, $CH_3OH$, $H_2CO$, CN, CS). The low enrichment in $^{13}C$ for $CH_3OH$ have already been predicted by Charnley *et al.* (2004) and observed few years latter toward various stars (Wirström *et al.* 2011). The $^{13}C$ fractionation of methanol reflects the fractionation of CO in the gas phase, confirming that methanol is very likely formed by protonation of CO on ice. In contrast, species showing a strong depletion in $^{13}C$ (c.$C_3H_2$, $C_2H$, $^{13}CCS$) are not well reproduced at all because they are linked to $C_3$ which is highly enriched in $^{13}C$.

For c-$C_3H_2$ there are two measurements, both leading to similar results. A slight depletion is observed for the asymmetric form, c-$C_3H_2$/c-$C^{13}CHCH$ equal to 61 ± 11 in TMC-1 (Yoshida et al. 2015) and 53 ± 16 in L483 (Agúndez et al. 2019), compared with the expected value of 34 taking into account the statistical factor of 2 for the two equivalent carbon atoms c-$C^{13}CHCH$ and c-$CCH^{13}CH$. A strong depletion is observed for the symmetric form with a ratio $C_3H_2$/c-$^{13}CCHCH$ equal to 310 ± 80 in TMC-1 and 458 ± 138 in L483. In our simulations, we obtain, for a chemical age between $3\times10^5$ and $7\times10^5$ years, $C_3H_2$/c-$C^{13}CHCH$ = 7 (instead of $61 \pm 11$ and $53 \pm 16$ observed) and $C_3H_2$/c-$^{13}CCHCH$ = 43 (instead of $310 \pm 80$ and $458 \pm 138$ observed). Obviously, our results are not in agreement with the observed isotope ratios even if the simulated c-$^{13}CCHCH$/ c-$C^{13}CHCH$ ratio itself (equal to 6 between $3\times10^5$ and $7\times10^5$ years) is relatively close the observed ones ($5.1 \pm 2.3$ in TMC-1 and $10.6 \pm 4.3$ in L483).

$C_2H$ and $C_2S$ are also linked to $C_3$ although less directly than c-$C_3H_2$. For both $^{13}CCH$ and $C^{13}CH$ as well as $^{13}CCS$, the simulations lead to $^{13}C$ enrichment levels that are far too high when compared with observations.

Another species that is potentially directly related to $C_3$ is methylacetylene ($CH_3CCH$). In the isotope network we did not include $CH_3CCH$ as this species is difficult to form in the gas-phase in our model. Here, the ionic pathway via $C_3H_3^+ + H_2$ is inefficient according to Lin et al. (2013a), while current desorption mechanisms do not allow enough surface formed $CH_3CCH$ to be liberated into the gas-phase despite its relatively large abundance on grains (formed from the hydrogenation of s-$C_3$). If we consider that $CH_3CCH$ forms exclusively from $C_3$ reactions (which is not certain), we expect that the $^{13}C$ enrichment of the $C_3$ skeleton to be preserved in species related to $C_3$. Then we will not be able to explain the observations for L483, as we should have large $^{13}C$ fractionation in the central position of the carbon skeleton (or for the mean of the three positions if there are carbon skeleton rearrangement for $CH_3CCH$ production from $C_3$ species).

Consequently, it is essential to examine the role of $C_3$ in $^{13}C$ fractionation. In cold molecular clouds, it is impossible to detect the $C_3$ molecule which possesses no dipole moment, while its abundance is too weak for its detection by direct absorption (Roueff *et al.* 2002). At higher temperatures, it is possible to populate excited bending vibrational levels which induce a dipole moment, therefore allowing $C_3$ to be detected in dense media (Cernicharo *et al.* 2000, Giesen *et al.* 2019). Very recently, $^{13}CCC$ and $C^{13}CC$ have been detected in SgBr2(M) (Giesen et al. 2019) with an average $^{12}C/^{13}C$ ratio equal to $20 \pm 4.2$, which is close to the elemental ratio for SgrB2 (SgrB2 is close to center of the Galaxy with a lower elemental $^{12}C/^{13}C$ ratio, (Milam et al. 2005)). Additionally, the $^{13}CCC/C^{13}CC$ ratio was found to be equal to $1.2 \pm 0.1$ instead of the statistically expected value of 2, showing some $^{13}C$ fractionation with the central position favored. The observations were made on a warm envelope, thus with a very different chemistry where desorption from ices plays a large role and where relatively high temperatures should limit isotopic fractionation. It should be noted that in addition to $C_3$ (Giesen et al. 2019), the molecules CO, c-$C_3H_2$ and $C_2H$ also display little or no $^{13}C$ fractionation in SgBr2(M) (Belloche

*et al.* 2013). Realistic chemical simulations of SgBr2(M) are complex as this region is made up of many fragments with a temperature up to 170 K in some parts, even if the region probed by $C_3$ detection is thought to be mainly the envelope with a colder temperature (the average temperature of the observations is estimated to 44 K). Moreover, such simulations would require a good knowledge of the history process of SgBr2(M) formation, as the history of dense core formation, prior to the warm-up phase, plays a large role in isotope fractionation if the compounds present on the grains are derived from species produced in the gas phase at low temperatures. However, despite all the uncertainties, some $^{13}C$ enrichment is expected in $C_3$ for SgBr2(M).

Considering the $^{13}C$ depletion for c-$C_3H_2$ and $C_2H$, $^{13}CCS$ in various molecular clouds, and the low $^{13}C$ fractionation for $CH_3CCH$ in L483 and $C_3$ in SgrB2(M), the importance of $C_3$ as a $^{13}C$ reservoir is questionable.

## 3.2 Comparison with $^{13}C$ previous studies of Roueff et al. (2015) and Colzi et al. (2020))

In Roueff et al. (2015), the model did not consider reactions on grains and the isotopic exchange reactions 1, 2, 3, 8-20 were not included, which induces significant differences when compared with our results, especially since there were no fractionation reactions of $C_3$. The model in Colzi et al. (2020) is much more similar to our model, including grain chemistry and considering some of the new fractionation reactions especially one for $C_3$ (reaction 8). They also consider that $C_3$ doesn't react with atomic oxygen and then $C_3$ reached a high abundance as in our nominal model. The models are different on 2 major points: Colzi et al. (2020) did not take into account non-equivalent carbons and did not take into account some important fractionation reactions (reactions 9, 11, 12, 13, 14, 15, 16, 17, 18, 19 and 20 of Table 2). A comparison for time dependence of the $^{12}C/^{13}C$ ratio given by our nominal model and the one from Colzi et al. (2020) is shown in Figure 2 for various key species (in both cases the parameters, such as density, temperature, initial conditions,… were identical). As in their study Colzi et al. (2020) have considered all carbons of a given species as equivalents, they implicitly sums up all the isotopologues of the same species (for $C_3$, $C_2H$ and $C_2S$) and we must do the same to be able to compare the results. The assumption that all carbons of a given species are equivalent is based on the assumption that all reactions proceed via full scrambling of carbon atoms which is not universally true, as they noted in their article, and is not coherent with the observations for $C_2H$, $C_2S$, c-$C_3H_2$ and $HC_3N$ (see Table 1).

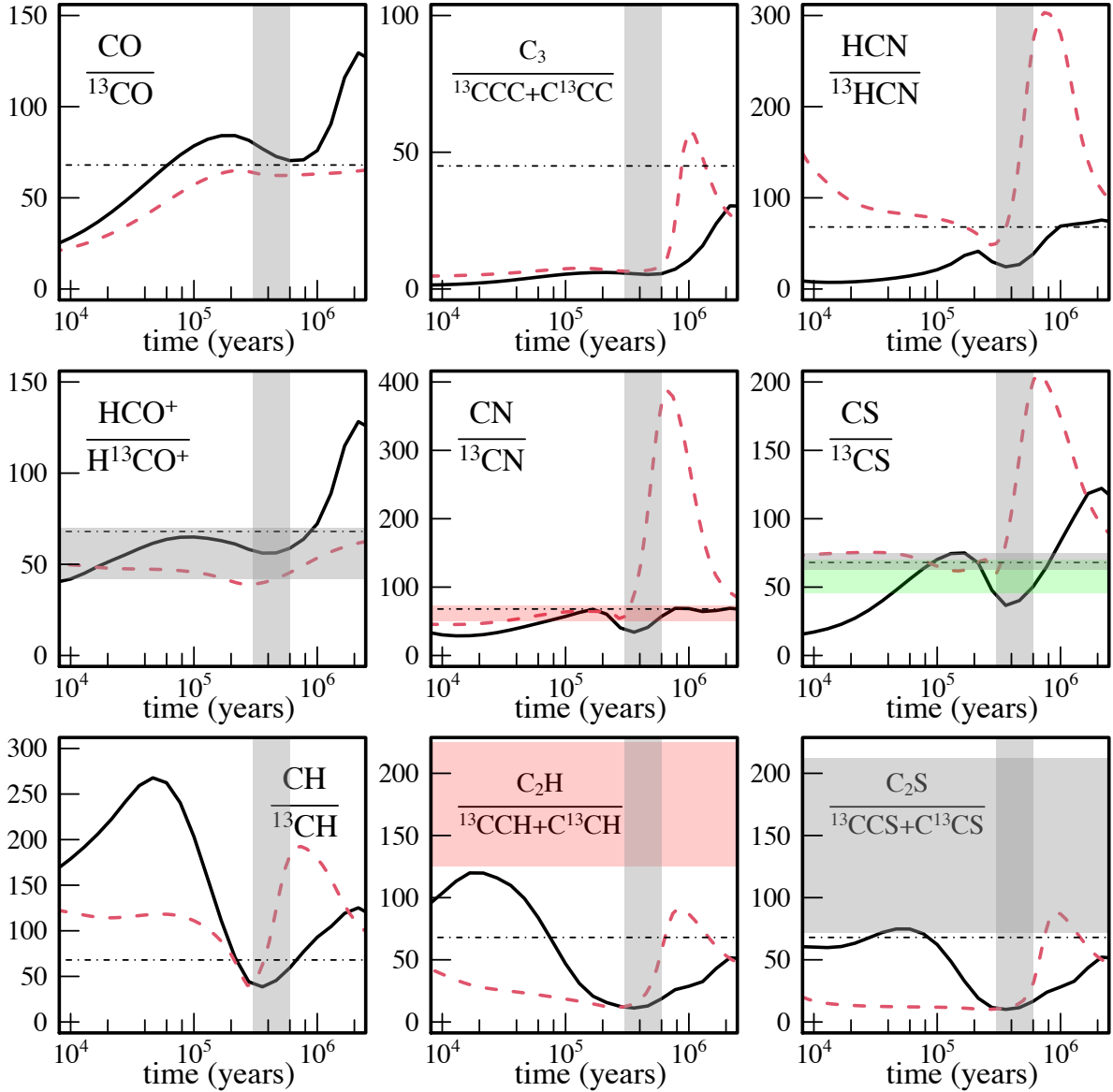

**Figure 2**: Gas phase species of $^{12}C/^{13}C$ ratio abundances of various species as a function of time predicted by our model (n(H$_2$) = 2×10$^4$ cm$^{-3}$, T = 10K) with a rate constant for the O + C$_3$ reaction equal to 1×10$^{-14}$ cm$^3$.molecule$^{-1}$.s$^{-1}$ (in continuous black line) and from Colzi et al. (2020) (red dashed line). The dashed horizontal lines represent the "cosmological" $^{12}C/^{13}C$ ratio equal to 68 (45 for the C$_3$/($^{13}$CCC+C$^{13}$CC) taking into account the statistical factor of 2 for the two equivalent carbon atoms). The observations in dense molecular clouds are reported in horizontal rectangles (including the uncertainties), in light grey for TMC-1(CP), in light red for L1527, in light blue for L1498 and in light green for L483 (see Table 1 for the references). The vertical grey rectangles represent the values given by the most probable chemical age for TMC-1(CP) and L483 given by the better agreement between calculations and observations for key species.

Up to few 10$^5$ years, both models leads to relatively similar $^{13}$C fractionation behavior for CO (and HCO$^+$) as well as C$_3$, both being the reservoirs of $^{13}$C. At the longest time the models diverge, C$_3$ not remaining a $^{13}$C reservoir in their model unlike ours, this behavior may be due to the fact that they did not take into account the most stable C$^{13}$CC isotopologue. For CN the curves are also similar for time below 2×10$^5$ years, the $^{13}$CN enrichment being due to reactions 4 and 10. For longer time, CN is mainly produced through the DR of HCNH$^+$ and then the $^{13}$CN fraction follows the fractionation of HCN. The fact that Colzi et al. (2020) did not include the reactions of fractionation in $^{13}$C for HCN (our reactions 11, 12 and 13) explains the differences. This shows the importance of having a chemical network as complete as possible. This is also the case for CS for which the reaction 15, not present in Colzi et al. (2020), induces notable $^{13}$CS fractionation. The comparison for C$_2$H and C$_2$S is particularly complicated because, in addition to the intrinsic differences in the chemical networks, considering the two carbons as equivalent and not taking into account reactions 16 and 17 in Colzi et al. (2020) induce many biases. It can be noted, however, that both models predict a significant enrichment in $^{13}$C for C$_2$H and C$_2$S for typical ages of dense clouds, an enrichment that is in contradiction with observations as already noted in the previous sections. Colzi et al. (2020) did not include C$_3$H$_{x=1-2}$ species in their model, but since these species are very strongly chemically bound to C$_3$, the high enrichment of C$_3$ in $^{13}$C obtained by Colzi et al. (2020) is expected to be found in these species, as in our model, contrary to observations. Colzi et al. (2020) highlighted the potential role of C$_3$ as a $^{13}$C reservoir but did not question the results induced by the high $^{13}$C enrichment of C$_3$ as they focused on CN, HCN and HNC, species with little chemical connection to C$_3$.

### 3.3 Role of the O + C$_3$ reaction

As there is no doubt that the reaction 8 presents no barrier in the entrance valley (Wakelam et al. 2009) leading to efficient fractionation, the differences between the models (our nominal model as well as the one of Colzi et al. (2020)) and the observations may arise from an overestimation of the C$_3$ abundance in current models. Among all the different reactions that can consume C$_3$, the O + C$_3$ reaction is known to play a critical role (Hickson *et al.* 2016b). An earlier theoretical study by Woon and Herbst (1996) found a small barrier in the entrance valley leading to a negligible rate constant at low temperature. The Woon and Herbst (1996) study forms the basis for the rate constants used in our model ($1\times10^{-14}$ cm$^3$.molecule$^{-1}$.s$^{-1}$ at 10 K, the results being the same when considering a smaller rate constant). However, we performed for this work additional calculations on the O + C$_3$ reaction, at several levels with various basis sets showing that the height of the barrier is very method and basis set dependent (presentation of these calculations is out the scope of this paper and will be presented in a future paper). Consequently, there is considerable doubt regarding the low temperature rate constants of this reaction. Interestingly, a related process, the O + propene reaction is also characterized by a small barrier in the entrance valley but with non-negligible experimental rate constants at low temperature (Sabbah *et al.* 2007, Zhang *et al.* 2007). It can be noted that a recent observational and modeling study of C$_3$H$_6$ and C$_2$H$_3$CHO in IRAS16293B (Manigand *et al.*

2020) shows clearly that a much better agreement between the model and the observations is obtained when $C_3$ reacts with oxygen atom. In order to test the effect of the $O + C_3$ reaction on our carbon fractionation model, we performed simulations where atomic oxygen is allowed to react with $C_3$. The results of these simulations are shown in Figure 3, where the rate constant for the $O + C_3$ reaction has been set to $2\times10^{-12}$ cm$^3$.molecule$^{-1}$.s$^{-1}$ (dashed red lines) and $2\times10^{-11}$ cm$^3$.molecule$^{-1}$.s$^{-1}$ (dotted blue line).

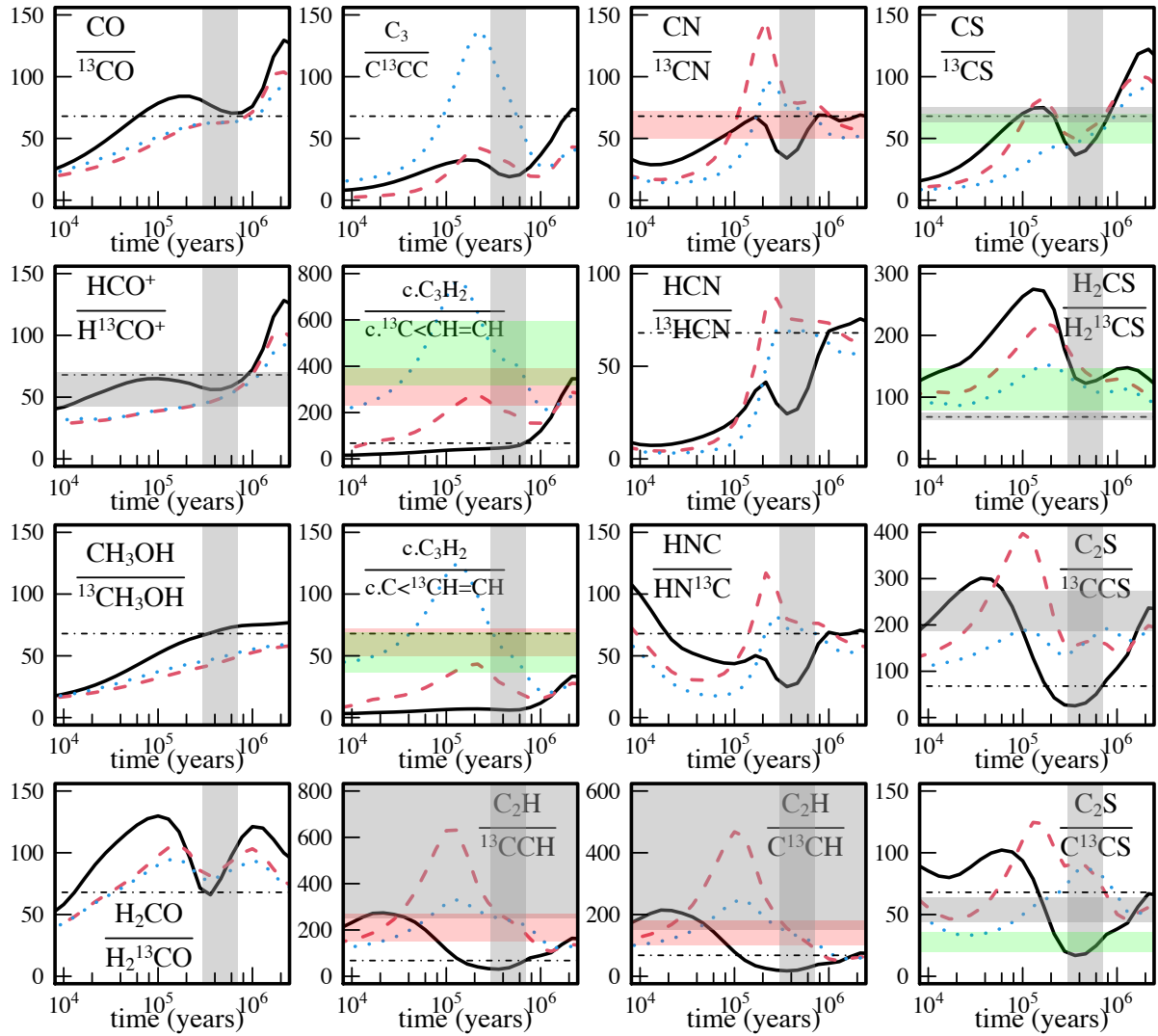

**Figure 3:** Gas phase species of $^{12}C/^{13}C$ ratio abundances of various species as a function of time predicted by our model (n(H$_2$) = $2\times10^4$ cm$^{-3}$, T = 10K) with a rate constant for the $O + C_3$ reaction equal to $1\times10^{-14}$ cm$^3$.molecule$^{-1}$.s$^{-1}$ (continuous black lines), equal to $2\times10^{-12}$ cm$^3$.molecule$^{-1}$.s$^{-1}$ (dashed red lines), and equal to $2\times10^{-11}$ cm$^3$.molecule$^{-1}$.s$^{-1}$ (dotted blue lines). The dashed horizontal lines on the bottom plots represent the "cosmological" $^{12}C/^{13}C$ ratio equal to 68 (c-C$_3$H$_2$/c-C$^{13}$CHCH taking into account the statistical factor of 2 for the two equivalent carbon atoms). The vertical grey rectangles represent the values given by the most probable chemical age for TMC-1(CP) and L483 given by the better agreement between calculations and observations for key species. The observations in dense molecular clouds are reported in horizontal rectangles (including the uncertainties), in light grey for TMC-1(CP), in light red for L1527, in light blue for L1498 and in light green for L483 (see Table 1 for the references).

When $C_3$ is allowed to react more rapidly with atomic oxygen, its abundance is significantly reduced as shown Figure 4 and it is much less enriched in $^{13}C$. Then, the compounds directly linked to $C_3$, especially c-$C_3H_2$, are poorly enriched in $^{13}C$ and even depleted for one isomer. For the $^{12}C/^{13}C$ ratio, a much better agreement between the model and observations for most of the species in dense clouds is obtained when $k(O + C_3)$ is set to an intermediate value of $2\times10^{-12}$ cm$^3$.molecule$^{-1}$.s$^{-1}$.

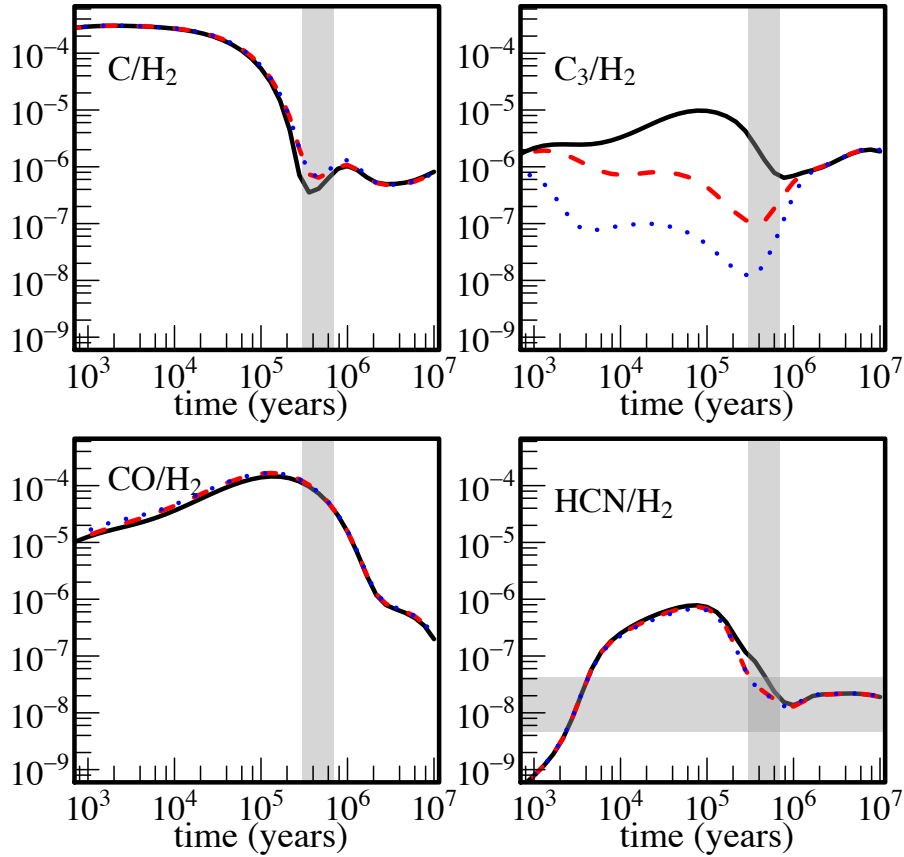

**Figure 4:** Gas phase C reservoirs (C, CO, $C_3$, HCN) as a function of time predicted by our model ($n(H_2) = 2\times10^4$ cm$^{-3}$, T = 10K) with a rate constant for the O + $C_3$ reaction equal to $1\times10^{-14}$ cm$^3$.molecule$^{-1}$.s$^{-1}$ (continuous black lines), equal to $2\times10^{-12}$ cm$^3$.molecule$^{-1}$.s$^{-1}$ (dashed red lines), and equal to $2\times10^{-11}$ cm$^3$.molecule$^{-1}$.s$^{-1}$ (dotted blue lines). The vertical grey rectangles represent the values given by the most probable chemical age given by the better agreement between calculations and observations for key species. The HCN observation in TMC-1 is reported in horizontal rectangles (including the uncertainties) (see Table 1 for the reference).

In this case, CO and the daughter species that derive from CO (s-$CH_3OH$, $CH_3OH$ but also s-$H_2CO$ and s-$CO_2$) are enriched in $^{13}C$ at times characteristic of dense cloud ages (around 3-$7\times10^5$ years at a density of $n(H_2) = 2\times10^4$ cm$^{-3}$). However, as the main reservoir of atomic carbon, CO and all the species that derive from CO (such as those formed on grains) inherit similar weak enrichment factors in good agreement with Charnley et al. (2004) and Wirström et al. (2011). The low enrichment of CO does, however, lead to $^{13}C$ depletion for all other species less correlated with CO. For a certain number of them, such as HCN and CS, certain fractionation reactions ($^{13}$C + HCN, $^{13}$C + CS) partly compensate for the depletion effect generated by CO enrichment. However, these processes involving atomic carbon have a limited effect because carbon is mainly locked up as CO at the typical ages of dense clouds, so its abundance in the gas phase is too low to induce significant fractionation. In addition, the secondary reservoirs of carbon such as $C_3$, HCN and CS have a non-negligible reactivity. Indeed, protonation reactions induce a destruction of $C_3$, HCN and CS because the protonated forms do not recycle back to the precursor species with 100% efficiency contrary to CO. Then, as $C_3$, HCN and CS have a non-negligible reactivity, they do not accumulate, in contrast to CO, and the fractionation produced between $10^4$ and $10^5$ years does not persist at later times, except for CO.

### 3.3.1 HCN

The slight $^{13}$C enrichment of HCN and HNC observed in some dense molecular clouds (Daniel et al. 2013, Magalhães et al. 2018) is relatively well reproduced by our model when $C_3$ reacts with oxygen atoms for short chemical ages ($2\times10^5$ years), such a short time being compatible with the chemical age for L1498 (Magalhães et al. 2018) and B1b (Daniel et al. 2013) given their high densities which accelerate chemistry. It should be noted that the determination of the abundance of the main isotopes ($H^{12}C^{14}N$, $H^{14}N\,^{12}C$) is difficult in particular due to the opacity of the lines.

### 3.3.2 CS

Unlike HCN and HNC, our model predicts a slight enrichment in $^{13}$C for CS due to the $^{13}$C + CS reaction. The comparison with the observations is rather confusing in this case because, depending on the measurements, CS is sometimes enriched and sometimes depleted in $^{13}$C. It should be noted that the $^{12}C^{32}S$ lines are generally optically thick and that the double isotope method is used to determine the level of $^{13}$C fractionation assuming a $^{32}S/^{34}S$ ratio equal to the solar ratio. This method seems credible for CS according to Loison et al. (2019) showing a low $C^{34}S$ enrichment in dense clouds for non-depleted sulfur, which becomes completely negligible when sulfur is depleted, as observed in dense clouds.

### 3.3.3 H$_2$CO

H$_2$CO is an interesting case showing variable $^{13}$C fractionation, the model being in good agreement with the observation showing a small $^{13}$C depletion for typical chemical age of dense molecular clouds. The variation in $^{13}$C fractionation is due to the fact that H$_2$CO is formed by two main pathways. Firstly, a gas phase pathway: O + CH$_3$, inducing $^{13}$C depletion as CH$_3$ is directly derived from atomic carbon depleted by CO enrichment. Second, a grain pathway by hydrogenation of s-CO showing an enrichment similar to that of methanol.

### 3.3.4 H$_2$CS

Contrary to H$_2$CO, the model leads to a substantial depletion for H$_2$CS which is not in agreement with the observations of Liszt and Ziurys (2012) showing no depletion. The depletion given by the model is due to the fact that H$_2$CS is produced essentially from reactions in the gas phase from CH$_4$ or CH$_3$, and as there are no known $^{13}$C enrichment reactions for CH$_4$, CH$_3$ and H$_2$CS, it is indeed expected that H$_2$CS is depleted in $^{13}$C due to CO enrichment. Future observations seem important to clarify this point as H$_2$CS may be used as a proxy for $^{13}$C fractionation of CH$_3$ and CH$_4$.

### 3.3.5 C$_2$H

For C$_2$H there are no direct fractionation reactions leading to notable $^{13}$C enrichment of C$_2$H. Then, when C$_3$ is allowed to react with oxygen atoms, the chemical pathway linking C$_3$ to C$_2$H become less important and C$_2$H become depleted in $^{13}$C that is more or less marked depending on whether they are related to partially enriched species. Overall, the depletion observed for C$_2$H is fairly well reproduced in the model. For C$_2$H the carbon atoms are not equivalent and the observed isotope fractionation depends on the carbon position. As suggested by Sakai et al. (2007) and Sakai et al. (2010), the different fractionation for non-equivalent carbons in our network is mainly due to reactions with atomic hydrogen (reaction 16). C$_2$H is not correlated to CO and, since there are no fractionation reactions either for these species or its main precursors (when C$_3$ is allowed to react with atomic oxygen, the various pathways connecting C$_3$ to C$_2$H involve only small fluxes), C$_2$H then shows a global depletion in $^{13}$C for both isomers. The reaction 16 favors the C$^{13}$CH isomer, but the energy difference is too low to induce a strong effect, the ratio given by the simulations being in good agreement with the observations of Sakai et al. (2010). As shown in Figure 5 the calculated ratio C$^{13}$CH/$^{13}$CCH is mostly independent of the rate constant for the O + C$_3$ reaction.

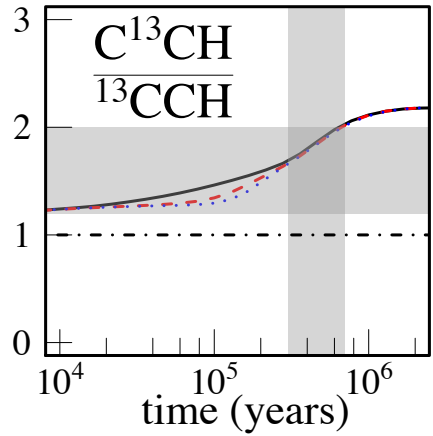

**Figure 5:** Gas phase $C^{13}CH/^{13}CCH$ ratio as a function of time predicted by our model ($n(H_2) = 2\times10^4$ cm$^{-3}$, T = 10K) with a rate constant for the O + C$_3$ reaction equal to $1\times10^{-14}$ cm$^3$.molecule$^{-1}$.s$^{-1}$ (continuous black lines), equal to $2\times10^{-12}$ cm$^3$.molecule$^{-1}$.s$^{-1}$ (dashed red lines), and equal to $2\times10^{-11}$ cm$^3$.molecule$^{-1}$.s$^{-1}$ (dotted blue lines). The vertical grey rectangles represent the values given by the most probable chemical age given by the better agreement between calculations and observations for key species. The observations for TMC-1(CP) is reported in horizontal light grey rectangle (see Table 1 for the reference).

### 3.3.6 C$_2$S

For C$_2$S, as for C$_2$H, there are no fractionation reactions leading to notable $^{13}$C fractionation of C$_2$S. Then, when C$_3$ is allowed to react with oxygen atoms, C$_2$S become depleted in $^{13}$C, with an observed depletion for $^{13}$CCS + C$^{13}$CS fairly well reproduced in the model. As for C$_2$H, the carbon atoms are not equivalent and the observed isotope fractionation depends on the carbon position for which the reaction 18 plays a major role, promoting the formation of the C$^{13}$CS isomer. As the energy difference between the two isomers is notably larger than in the case of C$_2$H (18 K instead of 8 K) the fractionation effect is larger reaching a C$^{13}$CS /$^{13}$CCS ratio equal to 4.2 ± 2.3, in TMC-1 (Sakai et al. 2007).

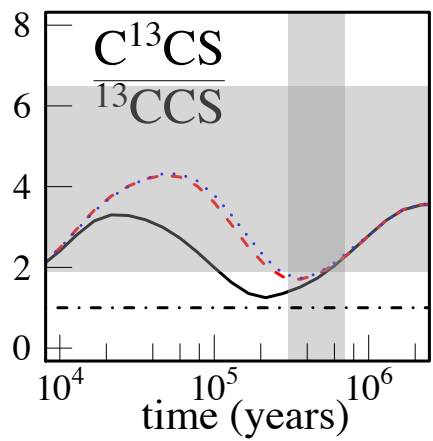

**Figure 6:** Gas phase $C^{13}CS/^{13}CCS$ ratio as a function of time predicted by our model ($n(H_2) = 2\times10^4$ cm$^{-3}$, T = 10K) with a rate constant for the O + C$_3$ reaction equal to $1\times10^{-14}$ cm$^3$.molecule$^{-1}$.s$^{-1}$ (continuous black lines), equal to $2\times10^{-12}$ cm$^3$.molecule$^{-1}$.s$^{-1}$ (dashed red lines), and equal to $2\times10^{-11}$ cm$^3$.molecule$^{-1}$.s$^{-1}$ (dotted blue lines). The vertical grey rectangles represent the values given by the most probable chemical age given by the better agreement

between calculations and observations for key species. The observations for TMC-1(CP) is reported in horizontal light grey rectangle (see Table 1 for the reference).

The ratio given by the simulations is not in good agreement with the observations, the ratio being underestimated. It can be noted that, as shown in Figure 6, the calculated ratio is only slightly dependent of the O + $C_3$ rate constant. One reason for the relative disagreement is due to the fact that even if the H-atom reaction favors the formation of the $C^{13}CH$ isomer, the main source of $C_2S$ is the S + $C_2H$ reaction and as the ratio $C^{13}CH/^{13}CCH$ is above 1 in our model, in agreement with observations, the S + $C^{13}CH \rightarrow {}^{13}CCS + H$ reaction has a higher flux than the S + $^{13}CCH \rightarrow C^{13}CS + H$ reaction. The S + $C_2H$ reaction is an efficient $C_2S$ pathway production as it cannot lead to CH + CS which is endothermic. The S + $^{13}CCS \rightarrow C^{13}CS + S$ reaction, proposed by Sakai et al. (2010), leads to the very exothermic CS + CS exit channel and cannot play an important role in $^{13}C$ fractionation. However, for shorter times, around $10^5$ years, the $C^{13}CS/^{13}CCS$ is in better agreement with observations due to the CH + CS → H + CCS reaction, as CS is slightly enriched in $^{13}C$ and CH slightly depleted. The disagreement of the simulations and the observations at typical chemical age of dense molecular clouds may be due to the overestimation of the S + $C_2H$ pathway (if the rate constant is lower than estimated) associated to the underestimation of the CH + CS one. It should be noted that our model for isotopes, as well as our full model, underestimates the $C_2S$ abundance and therefore some $C_2S$ production routes could be missing which might lead to a different level of $^{13}C$ fractionation.

### 3.3.7 c-$C_3H_2$

c-$C_3H_2$ is a more complex case. c-$C_3H_2$ has also non-equivalent carbons atoms (see appendix for the drawing of the two isotopologues). In our model, c-$C_3H_2$ is essentially produced through the DR of c-$C_3H_3^+$ and l-$C_3H_3^+$, both formed from $C_3$ (Loison *et al.* 2017). As we consider that the CCC skeleton is scrambled in these reactions, which is a reasonable assumption considering the very large exothermicity of these reactions, we do not expect the enrichment of the central carbon atom of $C_3$ to be transferred to c-$C_3H_2$ as an enrichment of a single carbon atom. However, the model, in fairly good agreement with the observations as shown in Figure 7, shows a $^{13}C$ enrichment of c-$C_3H_2$ for the two equivalent hydrogen bearing carbon atoms and a depletion for the other. This effect is in fact due to the reaction 17 which has a strong effect due to the large exothermicity.

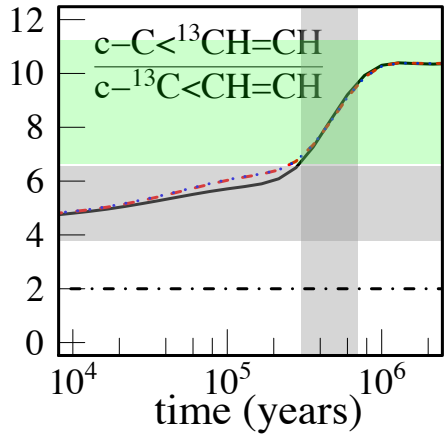

**Figure 7:** Gas phase c-C<$C^{13}$HCH/c-$^{13}$C<CHCH ratio as a function of time predicted by our model (n($H_2$) = 2×$10^4$ $cm^{-3}$, T = 10K) with a rate constant for the O + $C_3$ reaction equal to 1×$10^{-14}$ $cm^3.molecule^{-1}.s^{-1}$ (continuous black lines), equal to 2×$10^{-12}$ $cm^3.molecule^{-1}.s^{-1}$ (dashed red lines), and equal to 2×$10^{-11}$ $cm^3.molecule^{-1}.s^{-1}$ (dotted blue lines). The vertical grey rectangles represent the values given by the most probable chemical age for TMC-1(CP) and L483 given by the better agreement between calculations and observations for key species. The observations for TMC-1(CP) is reported in horizontal light grey rectangle and in light green rectangle for L483 (see Table 1 for the reference).

The c-C<$C^{13}$HCH/c-$^{13}$C<CHCH ratio is not dependent on the value of the rate constant for the O + $C_3$ reaction due to the importance of the reaction **17**. However, the total abundance relative to $H_2$ is highly dependent on the rate of the O + $C_3$ reaction as well as the global fractionation level. As shown in Figure 2, the global fractionation observations cannot be reproduced if the O + $C_3$ reaction rate constant is lower than a few $10^{-12}$ $cm^3.molecule^{-1}.s^{-1}$. It is interesting to note that when $C_3$ reacts with oxygen atom, its abundance is much lower, as is the abundance of c-$C_3H_2$. The model reproduces the observed total abundance of c-$C_3H_2$ for relatively late molecular cloud ages only (above 6×$10^5$ years for n($H_2$) = 2×$10^4$ $cm^{-3}$ as shown in Figure 8). For these ages the concentration of atomic oxygen is low, with atomic oxygen being mainly depleted from the gas-phase and transformed into $H_2O$ on grains. Thus, the consumption of c-$C_3H_2$ is low because the reaction O + c-$C_3H_2$ is the main pathway for c-$C_3H_2$ loss, the production of c-$C_3H_2$ being relatively important due to the dissociation of CO by cosmic rays and the chemistry generated by the desorption of $CH_4$ produced on the grains. This allows the abundance of c-$C_3H_2$ to reach levels similar to the observations but only for chemical ages above 6×$10^5$ years.

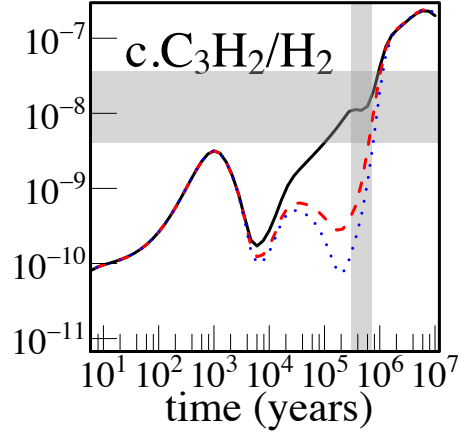

**Figure 8:** Gas phase c-$C_3H_2$ abundance relative to $H_2$ as a function of time predicted by our model ($n(H_2) = 2\times10^4$ $cm^{-3}$, T = 10K) with a rate constant for the O + $C_3$ reaction equal to $1\times10^{-14}$ $cm^3.molecule^{-1}.s^{-1}$ (continuous black lines), equal to $2\times10^{-12}$ $cm^3.molecule^{-1}.s^{-1}$ (dashed red lines), and equal to $2\times10^{-11}$ $cm^3.molecule^{-1}.s^{-1}$ (dotted blue lines). The vertical grey rectangles represent the values given by the most probable chemical age given by the better agreement between calculations and observations for key species. The observation in TMC-1 is reported in horizontal rectangles (including the uncertainties) (see Table 1 for the reference).

### 3.3.8 $HC_3N$

$HC_3N$ has formation pathways, with some leading to $HC_3N$ having low variable $^{13}C$ depletion levels for each carbon position (such as the CN + $C_2H_2$ reaction due to $C_2H_2$ depletion, $C_2H_2$ being depleted because main $C_2H_2$ production is issued from $C^+$ + CH reaction, both $C^+$ and CH being depleted in $^{13}C$), while some enrich $HC_3N$ in $^{13}C$ due to its connection with $C_3$ chemistry (through the reactions N + $C_3H_3$, N + t-$C_3H_2$, …). In figure 9 we present the results of our model (assuming some arbitrary choices as described in section 2.2.1) and the comparison with the observations leads to a globally good agreement if $C_3$ is allowed to react with oxygen atoms. As for c-$C_3H_2$ and $C_2H$, and despite the uncertainties, the high averaged enrichment of $HC_3N$ to $^{13}C$ when $C_3$ does not react with oxygen atoms is inconsistent with the observations.

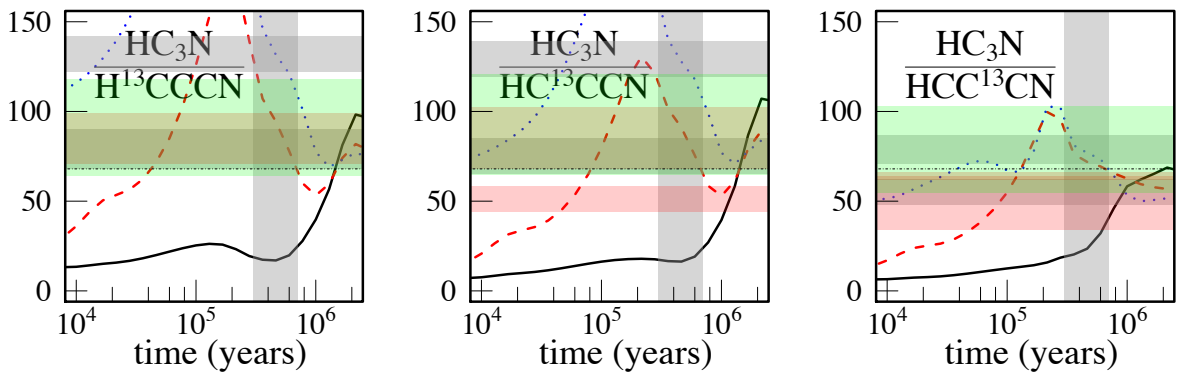

**Figure 9:** Gas phase $HC_3N/H^{13}CCCN$, $HC_3N/HC^{13}CCN$ and $HC_3N/HCC^{13}CN$, ratios as a function of time predicted by our model ($n(H_2) = 2\times10^4$ $cm^{-3}$, T = 10K) with a rate constant for the O + $C_3$ reaction equal to $1\times10^-$

$^{14}$ cm$^3$.molecule$^{-1}$.s$^{-1}$ (continuous black lines), equal to 2×10$^{-12}$ cm$^3$.molecule$^{-1}$.s$^{-1}$ (dashed red lines), and equal to 2×10$^{-11}$ cm$^3$.molecule$^{-1}$.s$^{-1}$ (dotted blue lines). The vertical grey rectangles represent the values given by the most probable chemical age for TMC-1(CP) and L483 given by the better agreement between calculations and observations for key species. The observations for TMC-1(CP) is reported in horizontal light grey rectangle, in light red rectangle for L1527 and in light green rectangle for L483 (see Table 1 for the reference). For TMC-1 and L1527 different observations exists leading to different isotopic ratios.

Considering the various effects, the observations with variable fractionation around the cosmological value (68), with H$^{13}$CCN and HC$^{13}$CN slightly depleted and HCC$^{13}$CN very close to the elemental value, are consistent with the HC$_3$N formation pathways in our model but only when C$_3$ reacts with atomic oxygen.

### 3.4 H$^{13}$CN/HC$^{15}$N, HN$^{13}$C/H$^{15}$NC and $^{13}$CN/C$^{15}$N

As the transitions of the main isotopologues of CN, HCN and HNC are often optically thick, most of the observational values for the ratios HCN/HC$^{15}$N, HNC/H$^{15}$NC, and some for CN/C$^{15}$N, are deduced from the ratios of the minor isotopologues H$^{13}$CN/HC$^{15}$N, HN$^{13}$C/H$^{15}$NC and $^{13}$CN/C$^{15}$N multiplied by a fixed local interstellar medium $^{12}$C/$^{13}$C value, usually taken to be 68 from Milam et al. (2005) and (Adande & Ziurys 2012). As suggested by Roueff et al. (2015) and shown in more detail here, carbon chemistry induces some $^{13}$C fractionation, leading to deviations from the elemental ratio of 6.49 (441/68) for H$^{13}$CN/HC$^{15}$N, HN$^{13}$C/HNC$^{15}$ and $^{13}$CN/C$^{15}$N (it should be noted that the elemental $^{14}$N/$^{15}$N ratio may be smaller than the usual 441 value as suggested by Romano et al. (2017), and could be as small as 330). In Figure 10 we show the various $^{13}$C/$^{15}$N ratios given by our model which includes also the $^{15}$N fractionation reactions (Loison et al. 2018) but neglecting any N$_2$ self-shielding effects of photodissociation (Furuya et al. 2018, Furuya & Aikawa 2018).

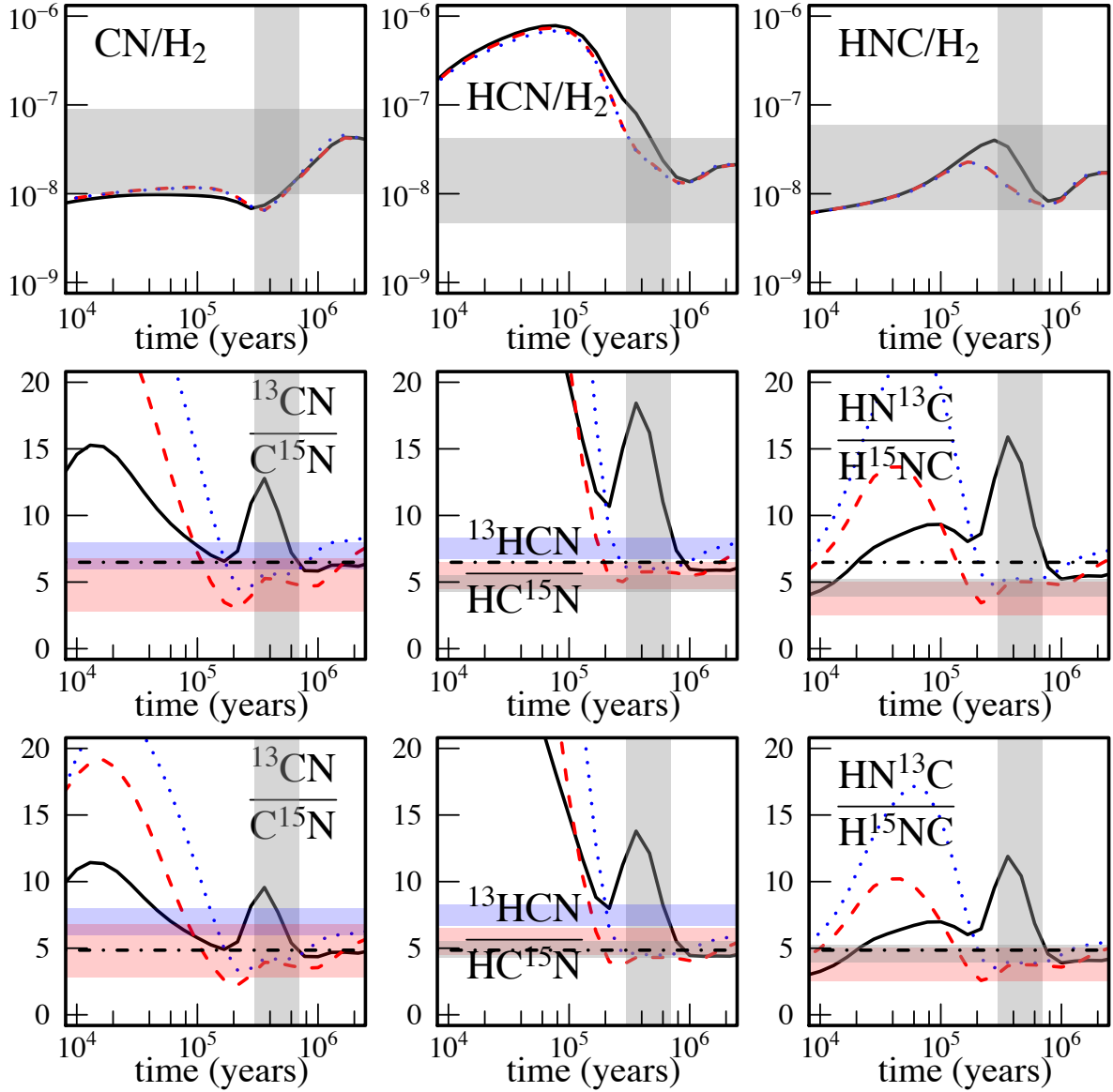

**Figure 10:** Gas phase species abundances (relative to $H_2$) of HCN, HNC, CN (top) and their $^{13}C/^{15}N$ ratio (medium and bottom, medium is for an elemental $^{13}C/^{15}N$ ratio equal to 441/68, scaled to 330/68 on the bottom plots) studied in this work as a function of time predicted by our model ($n(H_2) = 2\times10^4$ cm$^{-3}$, T = 10K) with a rate constant for the O + $C_3$ reaction equal to $1\times10^{-14}$ cm$^3$.molecule$^{-1}$.s$^{-1}$ (continuous black lines), equal to $2\times10^{-12}$ cm$^3$.molecule$^{-1}$.s$^{-1}$ (dashed red lines), and equal to $2\times10^{-11}$ cm$^3$.molecule$^{-1}$.s$^{-1}$ (dotted blue lines). The vertical grey rectangles represent the values given by the most probable chemical age for TMC-1(CP) and L483 given by the better agreement between calculations and observations for key species. The observations for TMC-1(CP) is reported in horizontal light grey rectangle, in light red rectangle for L1527 and in light blue for B1 (see Table 3 for the reference). The dashed horizontal lines on the medium plots represent the "cosmological" $^{13}C/^{15}N$ ratio equal to 441/68, scaled to 330/68 on the bottom plots.

The amount of the main isotopologues for HCN, HNC and CN are almost independent of the value of the O + $C_3$ rate constant. However, $^{13}C/^{15}N$ ratios are incompatible with observations in the nominal version of the model where $C_3$ does not react with oxygen atoms. Our model leads to H$^{13}$CN/HC$^{15}$N, HN$^{13}$C/HNC$^{15}$ and $^{13}$CN/C$^{15}$N ratios highly dependent on the cloud age due to variable $^{13}$C fractionation. For molecular clouds of the TMC-1 type (assumed to be best represented with a chemical age around 3-7×10$^5$ years here), our model leads to a H$^{13}$CN/HC$^{15}$N and HN$^{13}$C/HNC$^{15}$ ratios close to 5 and a $^{13}$CN/C$^{15}$N ratio close to 4. These results can be compared with the observations presented in Table 3, showing a relatively good agreement considering the uncertainties. It should be noted that the observations have been analyzed with various methods (LTE, LVG) and taking or not the opacity into account. Then the comparison between individual observations and also the comparison between observations and the model have to be taken with great care. A comparison with the results of Roueff et al. (2015) is complex because the global networks are very different and the chemistry of interstellar grains and C$_3$ enrichment are not included in Roueff et al. (2015). Nevertheless, it is clear that there are various efficient $^{13}$C fractionation reactions and that the double isotope method should be used carefully to determine $^{15}$N fractionation levels. The comparison with Colzi et al. (2020) is not possible as they do not consider $^{15}$N fractionation in their model.

**Table 3**: Observations for $^{13}$C/$^{15}$N ratio for CN, HCN and HNC in dense molecular clouds.

| species | ratio | Cloud | references |
|---|---|---|---|
| $^{13}$CN/C$^{15}$N | 7.0(1.0) | L1498 | (Hily-Blant *et al.* 2013) |
| | 7.5(1.0) | L1544 | (Hily-Blant et al. 2013) |
| | 4.8(2.0) | B1 | (Daniel et al. 2013), large C$^{15}$N uncertainty |
| | 6.2(1.3) | OMC-2 | (Kahane et al. 2018) |
| H$^{13}$CN/HC$^{15}$N | 2.7(0.4) | L1521E | (Ikeda *et al.* 2002) (average of the 4 positions) |
| | 4.9(0.6) | TMC-1 | (Ikeda et al. 2002) |
| | 9.5(2.0) | L1498 | (Ikeda et al. 2002) |
| | 7.5(0.8) | L1498 | (Magalhães et al. 2018) |
| | 5.5(1.0) | B1 | (Daniel et al. 2013) |
| | 5.6(0.8) | OMC-2 | (Kahane et al. 2018) |
| HN$^{13}$C/H$^{15}$NC | 4.59(0.64) | TMC1(CP) | (Liszt & Ziurys 2012) |
| | 7.56(1.44) | TMC1(NH$_3$) | (Liszt & Ziurys 2012) |
| | 4.78(1.36) | L1527 | (Liszt & Ziurys 2012) |
| | 3.75(1.25) | B1 | (Daniel et al. 2013) |
| | 6.0(0.8) | OMC-2 | (Kahane et al. 2018) |

## 3.5 Grain species

As there are several $^{13}$C enrichment reactions for different species, particularly for CO, and there is no reverse reaction producing a surplus of $^{13}$C carbon atoms, the carbon atoms in the gas phase are therefore depleted in $^{13}$C. Even though most of the carbon is converted to CO, and then to s-CO and its derivatives s-CH$_3$OH and s-CO$_2$, a significant part of the atomic carbon that depletes onto grains is partially converted to s-CH$_4$ in our model. The strong depletion of $^{13}$C in atomic carbon in the gas phase is thus retained in s-CH$_4$ as shown Figure 11.

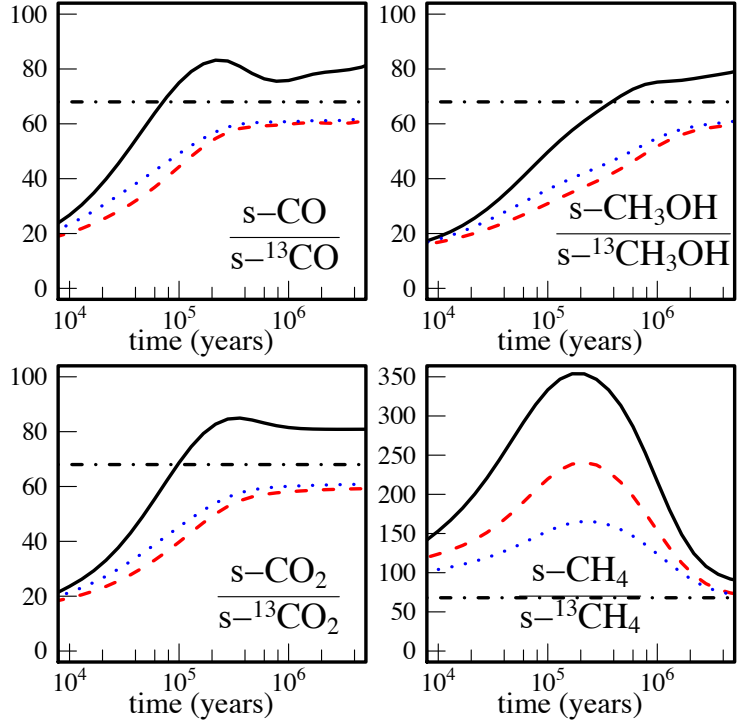

**Figure 11:** Gas phase species of $^{12}C/^{13}C$ ratio abundances of various species on grain as a function of time predicted by our model (n(H$_2$) = 2×10$^4$ cm$^{-3}$, T = 10K) with a rate constant for the O + C$_3$ reaction equal to 1×10$^{-14}$ cm$^3$.molecule$^{-1}$.s$^{-1}$ (continuous black lines), equal to 2×10$^{-12}$ cm$^3$.molecule$^{-1}$.s$^{-1}$ (dashed red lines), equal to 2×10$^{-11}$ cm$^3$.molecule$^{-1}$.s$^{-1}$ (dotted blue lines). The dashed horizontal lines on the bottom plots represent the "cosmological" $^{12}C/^{13}C$ ratio equal to 68.

Over a very long period of time this depletion is compensated by the fact that s-CH$_3$OH is partially transformed into s-CH$_4$ through photodissociation. If we consider that the composition of comets reflects the composition of the grains at the time of the formation of the protostellar disc, then this corresponds to ages around 10$^6$ years where CO and methanol have similar abundances. In this case the s-CH$_4$ present on the grains is significantly depleted in $^{13}$C. One could therefore expect the CH$_4$ observed in planetary atmospheres such as the one of Titan to show a similar depletion if the CH$_4$ present on Titan is primordial. However, the CH$_4$ on Titan shows a $^{12}C/^{13}C$ ratio equal to 91.1 ± 1.4 (Niemann *et al.* 2010) very similar to the $^{13}$C species detected in the comets (Bockelée-Morvan *et al.* 2015) or terrestrial ratio (Wilson 1999) which seems to indicate a reprocessing of the matter between the collapse of the molecular cloud, leading to the formation of the protostellar disc, and the formation of the planets. Alternatively, this may also indicate that carbon atoms are not mainly transformed into s-CH$_4$ by reactions with the highly mobile H and H$_2$ species on grain surfaces, whose reactions with atomic carbon are barrier-free (Harding *et al.* 1993, Krasnokutski *et al.* 2016). In the current model atomic carbon reacts primarily with water (Hickson *et al.* 2016a) leading to s-CH$_3$OH rather than s-CH$_4$ but some s-CH$_4$ is produced when carbon atom stick on Ice near s-CO$_2$, s-CH$_4$, s-C$_2$H$_6$ as we do not consider that the carbon atom will react on ice with these molecules by comparison with the gas phase reactivity.

## 4. Conclusion

Our new model presented in this study for coupled fractionation of $^{13}$C, $^{15}$N, $^{18}$O and $^{34}$S represents a significant improvement over previous models from Roueff et al. (2015) and Colzi et al. (2020). In particular, it shows the importance of using a network as complete as possible and also that the order on non-equivalent carbon atoms need to be tracked. Our results lead to various, in some cases notable, $^{13}$C fractionation effects due to efficient carbon chemistry of CO, and of $C_3$ when this species is assumed not to react with atomic oxygen. Due to the importance of $C_3$ as a reservoir of $^{13}$C, our study highlights the critical role of the O + $C_3$ reaction, with the $^{13}$C fractionation simulations of the nominal model which employ a low rate constant for this process, being largely incompatible with observations. Moreover, when $C_3$ doesn't react with oxygen atom, the nominal model strongly overestimate the $CH_3CCH$ abundance in the envelope of IRAS16293 (Andron *et al.* 2018), very likely because the model overestimates the $C_3$ abundance. Additionally, a recent observational and modeling study of $C_3H_6$ and $C_2H_3CHO$ in IRAS16293B (Manigand et al. 2020) shows clearly that a much better agreement between the model and the observations is obtained when $C_3$ reacts with oxygen atom. On the other hand, a rapid reaction of $C_3$ with atomic oxygen and the non-reactivity of $C_3H_3^+$ with $H_2$ as shown by Lin *et al.* (2013b), lead to an underestimation by several orders of magnitude of the concentrations of $CH_3CCH$ and $C_3H_6$ in the dense cloud (Hickson et al. 2016b). An experimental study of the O + $C_3$ rate constant is clearly required to resolve this issue.

The variable $^{13}$C fractionation given by our model, in quite good agreement with observations in dense molecular clouds, precludes the use of a fixed $^{12}$C/$^{13}$C ratio to determine accurately nitrogen fractionation using the double isotopes method.

The very low $^{13}$C fractionation observed in the solar system suggests a significant transformation of matter after the dark cloud phase. Indeed, our model predicts that some molecules should show significant fractionation levels at the end of this phase such as s-$CH_4$ (depleted in $^{13}$C) and s-$C_3H_x$ (enriched in $^{13}$C), levels that are not observed in comets, nor in planetary atmospheres. On the other hand, if the matter is heavily reprocessed, these species could be derived from CO and its derivatives (s-CO, s-$CH_3OH$, s-$CO_2$) showing only little fractionation since CO and its derivatives represent more than 90% of the total amount of carbon. Indeed, if the atomic carbon sticking on grains reacts with other species already on the grains, mainly water (Hickson et al. 2016a) and CO, this C atoms sticking will lead to very little

s-$CH_4$ and the $CH_4$ present in the grains would be derived from the photodissociation of s-$CH_3OH$ and would thus possess the same isotopic fractionation derived from the fractionation of CO (Wirström et al. 2011).

This work was supported by the program ˝Physique et Chimie du Milieu Interstellaire ˝ (PCMI) funded by CNRS and CNES. VW researches are funded by the ERC Starting Grant (3DICE, grant agreement 336474). We would like to warmly thank Laura Colzi for sharing her results with us, which allowed a fine comparison between the models.

**Data availability :** Data available on request.

## Appendix:

### $^{13}C + HCN$ and $^{13}C + HNC$:

For $^{13}C + HCN$ and $^{13}C + HNC$ reactions we have extended our previous study (Loison & Hickson 2015) showing that the carbon atom exchange is possible through two different $^{13}C$ attack pathways leading to $H^{13}CCN$ and $HC^{13}CN$ as shown on Figure 12. Although isotope exchange involves a reaction path involving several intermediates, the low energies of the TSs involved are probably compatible with our simplified isotope exchange treatment.

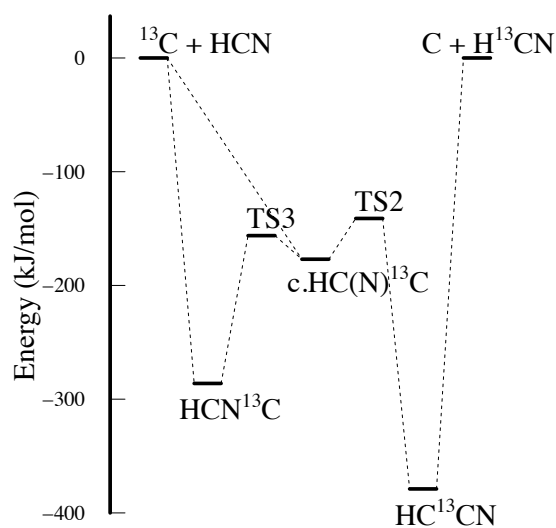

**Figure 12**: Partial potential energy diagram for the $^{13}C + HCN$ relevant for carbon atom exchange calculated at the M06-2X/AVTZ level including ZPE. The TS2 and TS3 correspond to the notation of our previous article (Loison & Hickson 2015).

In the case of $^{13}C + HNC$, the bimolecular exit channel $^{13}C + HCN$ limits the isotope exchange (however there is no pathway leading to $C + H^{13}CN$).

### $^{13}C + HCNH^+$:

For the $^{13}C + HCNH^+$ reaction we have performed DFT calculations showing that carbon atom exchange is possible through two different $^{13}C$ attack pathways leading to $H^{13}CCNH^+$ and $HC^{13}CNH^+$ as shown in Figure 13. Since isotope exchange involves several intermediates, with TS close to the entrance energy, our simplified treatment may overestimate the rate of exchange compared to back-dissociation.

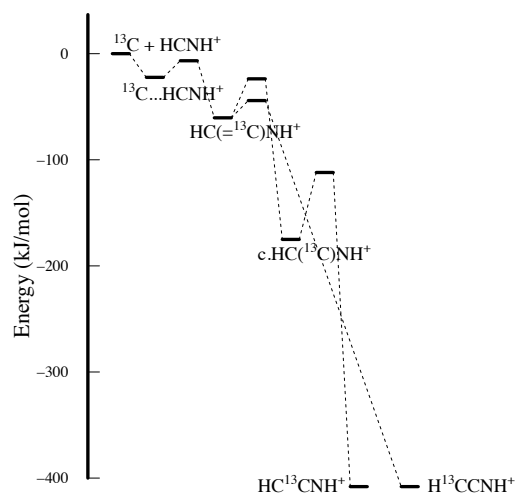

**Figure 13**: Partial potential energy diagram for the $^{13}C$ + HCNH$^+$ relevant for carbon atom exchange calculated at the M06-2X/AVTZ level including ZPE.

**$^{13}C$ + HC$_3$N:**

The $^{13}C$ + HC$_3$N reaction is particularly complex. This reaction has been studied theoretically by Li et al. (2006). According to this study, the additions on the different carbon atoms are barrier-free with however unknown relative proportions. As the exit channel on the triplet surface for this reaction, the H + C$_4$N pathway, is not very exothermic, some fractionation is expected. However, the precise mechanism for fractionation is very complex in that case as it involves C$_4$ skeleton rearrangements on the full surface for the various intermediates, which are produced in unknown relative proportions in the C + HC$_3$N reaction. In addition, spin-orbit coupling can promote the formation of HCN + C$_3$ which was neglected in the study by Li et al.

**$^{13}C$ + CS**

For the $^{13}C$ + CS reaction we performed new theoretical calculations at the M06-2X/AVTZ level using Gaussian and the RCCSD(T)-F12/AVTZ level using Molpro software. found no barrier in the entrance $^3\Sigma^-$ valley (linear approach) at the RCCSD(T)-F12/AVTZ level as shown in Figure 14. Moreover, in contrast to CCO, the transition state from $^{13}CCS$ toward S$^{13}CC$ is located well below the entrance level, making $^{13}C$ exchange possible. Additionally, carbon attack on the sulfur atom of CS presents a submerged barrier which also leads to $^{13}C$ exchange (not shown on Figure 14). Consequently, there is little doubt that the $^{13}C$ + CS reaction leads to $^{13}C$ exchange.

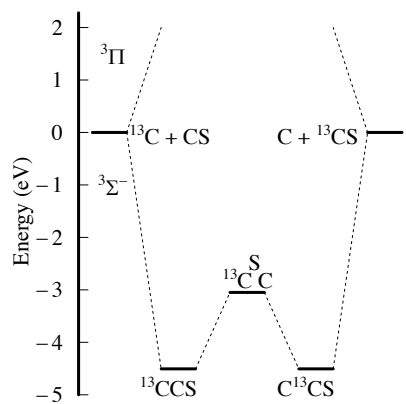

**Figure 14**: Potential energy diagram for the $^{13}$C + CS reaction. calculated at the CCSD(T)/AVTZ and M06-2X/AVTZ levels.

## H + c-C<$^{13}$CHCH/c-$^{13}$C<CHCH:

c-C<$^{13}$CHCH:                        c-$^{13}$C<CHCH:

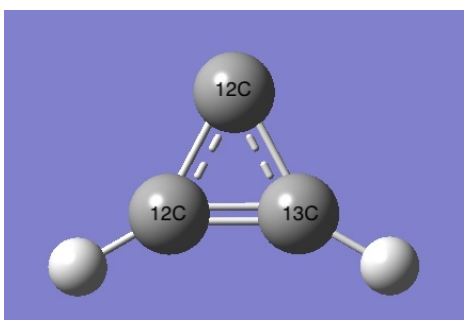 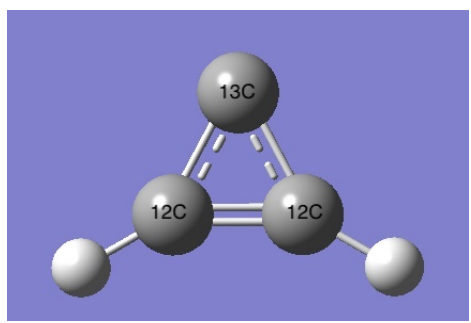

**Figure 15**: Drawing of the $^{13}$C isotopologues of c-C$_3$H$_2$

For the reaction of atomic hydrogen with cyclic c-C$_3$H$_2$ (with one $^{13}$C atom) reaction we use the work of (Nguyen *et al.* 2001b, Nguyen *et al.* 2001a) showing the possibility of isomerization through H atom addition. As this mechanism is direct and as there is no bimolecular exit channel, this reaction can be an efficient way to favor c-C<$^{13}$CHCH versus c-$^{13}$C<CHCH. Nguyen et al found a very small barrier at the RCCSD(T)/6-311+G//B3LYP/6-311G level (with ZPE at B3LYP/6-311G level) equal to 0.4 kJ/mol which is much smaller than the uncertainty of the calculations (this can be estimated at around 5-10 kJ/mol with this level of theory). We continued these calculations at different levels of theory and found that using a more complete base (VQZ) the barrier became slightly submerged (-3 kJ/mol). We therefore consider that this reaction is possible at low temperature even though the uncertainties are quite large.